\documentclass[twocolumn,showpacs,preprintnumbers,amsmath,amssymb]{revtex4}
\usepackage{graphicx}
\usepackage{amssymb}
\usepackage[latin1]{inputenc}
\usepackage{dcolumn}
\usepackage{bm}
\usepackage{natbib,hyperref}

\begin{document}


\title{Transport in Selectively Magnetically Doped Topological Insulator Wires}

\author{Sergio Acero$^1$, Luis Brey$^2$, William Herrera$^3$ and Alfredo Levy Yeyati$^1$}
\affiliation{
$^1$Departamento de F\'{\i}sica Te\'orica de la Materia Condensada, Condensed Matter Physics Center (IFIMAC) \\ and Instituto Nicol\'as Cabrera, Universidad Aut\'onoma de Madrid, E-28049, Spain\\
$^2$ Instituto de Ciencia de Materiales de Madrid, (CSIC),
Cantoblanco, E-28049 Madrid, Spain \\
$^3$Departamento de F\'{\i}sica, Universidad Nacional de Colombia, Bogot\'a, Colombia}
\date{\today}


\begin{abstract}
We study the electronic and transport properties of a topological insulator nanowire including selective magnetic doping of its surfaces. We use a model which is appropriate to describe materials like Bi$_2$Se$_3$ within a {\bf k.p} approximation and consider nanowires with a rectangular geometry. Within this model the magnetic doping at the (111) surfaces induces a Zeeman field which opens a gap at the Dirac cones corresponding to the surface states. For obtaining the transport properties in a two terminal configuration we use a recursive Green function method based on a tight-binding model which is obtained by discretizing the original continuous model. For the case of uniform magnetization of two opposite nanowire (111) surfaces we show that the conductance can switch from a quantized value of $e^2/h$ (when the magnetizations are equal) to a very small value (when they are opposite). We also analyze the case of non-uniform magnetizations in which the Zeeman field on the two
  opposite surfaces change sign at the middle of the wire. For this case we find that conduction by resonant tunneling through a chiral state bound at the middle of the wire is possible. The resonant level position can be tuned by imposing an Aharonov-Bohm flux through the nanowire cross section.  
\end{abstract}
\maketitle

\section{Introduction}

In some narrow gap semiconductors containing heavy atoms, the strong spin-orbit coupling induces an inversion of the energy bands with respect their customary energetic ordering. These materials are named Topological Insulators (TI's) \cite{hasan_2010,Qi_2011,Ortmann_2015}.
When in contact with vacuum, the electronic bands should recover their natural energy ordering and as a result a two-dimensional electron gas appears at the surface of the TI.
These surface states are governed by a massless Dirac Hamiltonian and have a conical dispersion centered at a time reversal invariant point in momentum space.
Angle resolved photoemission spectroscopy experiments have established the occurrence of these conical surface states in 
some topological materials of the family of Bi$_2$Se$_3$ \cite{Xia_2009,Chen_2009}.

Because the Dirac cone is located at a time reversal symmetry  invariant point,  only perturbations that break this symmetry can open a gap in the surface states. Apart from external magnetic fields, two methods can be used for opening  gaps at the Dirac cone: doping the surface with magnetic impurities or 
putting in contact the TI surface with a ferromagnetic material that induces
an exchange field \cite{He_2015}.  In the Bi$_2$Se$_3$ family of materials, the effective Hamiltonian for the electronic states of the (111) surface (hereafter denoted $z$-surface) takes the form $\hbar v_F( s_x k_y-s_y k_x)$, where  $s_x$, $s_y$ and $s_z$ are the electron spin Pauli matrices.
An exchange field pointing in the $z$-direction opens a gap in the Dirac cone of the surface bands. This gapped state is a non trivial insulator and it is characterized by a half-integer {\it anomalous} Hall conductivity $\sigma _{xy}$=$\pm e^2/2h$, where the orientation of the exchange field determines the sign of the conductivity \cite{Yokoyama_2010,Garate_2010}.

In real samples a surface is unavoidably connected with other surfaces and this prevents the observation of the half-integer quantized anomalous Hall effect.  Experimental observation of the quantum Hall effect (QHE) requires the existence of edge channels.  Each edge channel crossing the Fermi energy sums a contribution $e^2/h$ to the Hall conductivity, and therefore only integer QHE is expected to be observed. Topological insulator slabs with same sign exchange fields on top and bottom $z$-surfaces support chiral edge states connecting opposite surfaces and permit the observation of  integer anomalous quantum Hall effect, 
$\sigma _{xy}$=$e^2/h$ \cite{Lee_2009,Vafek_2011,Brey_2014}. In other combinations of exchange fields or  for other surface orientations, the lack of current carrying chiral edge modes makes the observation of anomalous QHE impossible \cite{Brey_2014,Sitte_2012,Chu_2011}.

Within this context, the case of TI nanowires has recently generated much interest 
\cite{Peng_2010,Kong_2010}. 
In these systems the surface state contribution could be more easily extracted in transport experiments,
as demonstrated by the observation of conductance oscillations associated to an
Aharonov-Bohm flux piercing the nanowire \cite{Peng_2010,Cho_2015} and by magnetoresistance measurements \cite{Gooth_2014}. The possibility to induce a magnetization of the surface states of a TI nanowire by selective magnetic doping has generated a number of theoretical studies 
\cite{Chu_2011,Brey_2014,Siu_2015}. These studies, however, have been mainly restricted to translational invariant geometries. Additionally, there are several studies of transport properties in hybrid TI systems \cite{PhysRevB.81.115407,PhysRevB.82.115211,PhysRevLett.104.046403,PhysRevB.86.035151,PhysRevB.87.035432,
PhysRevB.90.235148,PhysRevLett.115.096802,PhysRevB.92.081303}, but mainly considering infinite slabs or edges instead of finite nanowires. 

\begin{figure}
\begin{minipage}{1.0\linewidth}
 \includegraphics[width=1.0\textwidth]{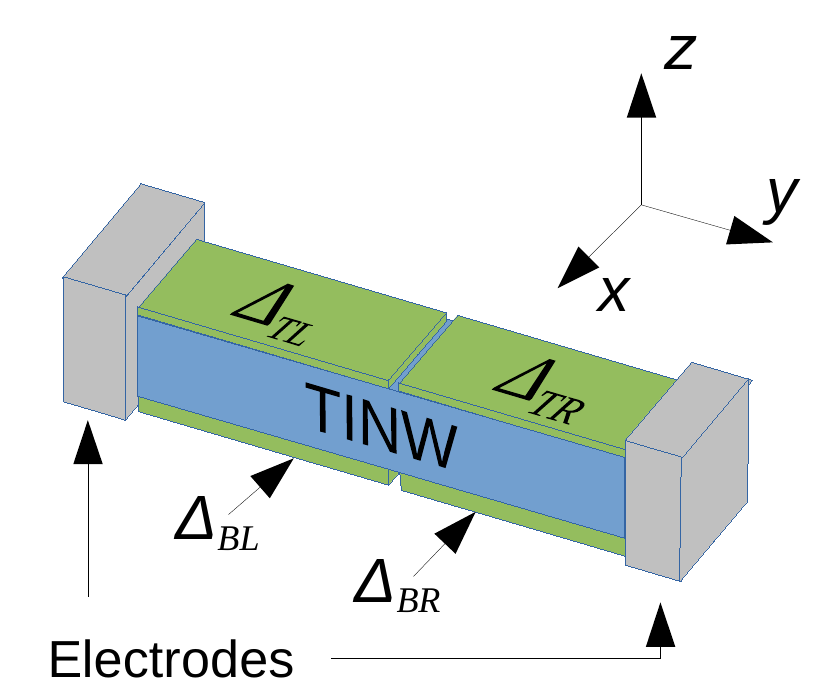}
\end{minipage}
 \caption{(Color online) Schematic representation of the topological insulator  nanowire (TINW) connected to metallic electrodes. The wire is directed along the $y$-direction, and on the top 
 and bottom $z$-surfaces there are applied exchanges fields, $\Delta _{TL}, \Delta _{BL}, \Delta _{TR}$ and $\Delta _{BR}$. }
 \label{TI-junction}
\end{figure}

In this work we study the transport properties of  selectively magnetically doped topological insulator quantum wires. 
We study a rectangular wire along the $\hat y$-direction connected to metallic leads, as schematically depicted in Fig. \ref{TI-junction}. 
Both, top (T) and bottom (B), $ z$-surfaces are divided in two equal left (L) and right (R) parts. 
At  each  sector there is  an exchange field applied on the surface pointing in the $\hat z$ direction. The wire is characterized 
by the exchange fields  ($\Delta _{TL},\Delta _{BL},\Delta _{TR},\Delta _{BR}$) and 
we study the electronic transport through wires with differently oriented  local exchange fields. In particular we focus in the transport between regions with opposite exchange fields.

Because the considered geometries are not translational invariant, we compute the conductance of the system using recursive Green functions techniques.
The electronic structure of the TI is calculated  with  a tight-binding (TB) Hamiltonian, that is  obtained by discretizing the ${\bf k} \cdot {\bf p}$ Hamiltonian that describe 
the band structure of Bi$_2$Se$_3$ near the  band gap energy \cite{Zhang_2009}.
The results obtained numerically with tight-binding calculations are discussed in terms of the two-dimensional Dirac-like effective Hamiltonian 
that control the mid gap states that appear at the surfaces. 

One of the main results of our work is to demonstrate that these devices can behave as a conductance switch controlled by the surface magnetization. Moreover, almost 100\% spin polarized currents could be generated for certain magnetic configurations.  On the other hand, we show that for an antisymmetric magnetic configuration a bound state confined in the middle of the wire appears. The system thus behave as a ``quantum dot" with a resonant level that can be controlled by a magnetic flux.  

The paper is organized in the following way: after the introduction we devote  section II to describe the model and the method for calculating the electronic and transport properties of the TI wire. In section III we present and discuss the main results. We conclude the paper in section IV with a summary of the main results.

\section{Model and Method}
\subsection{ $ {\bf k} \cdot {\bf p} $ model.}
In materials of the family of Bi$_2$Se$_3$  the low energy and long wavelength properties may be described quite accurately by the projection of the total Hamiltonian in  
a basis of 4 states, $|1>$=$|+,\uparrow>$, $|2>$=$-i|+,\uparrow>$, $|3>$=$|+,\downarrow>$ and $|4>$=$i |-,\downarrow>$, which are combinations of $p_z$ orbitals of Bi and Se with even $+$ and odd $-$ parities and spin up $\uparrow$ and down $\downarrow$.
The 4$\times$4 $\bf k \cdot \bf p$ Hamiltonian \cite{Zhang_2009} gets the form,
\begin{equation}
H^{3D}={\cal M }({\bf k}) \tau _z \otimes  I + A_1 k_z \tau _x \otimes s _z
+A_2 ( k_x \tau _x + k_y \tau _y) \otimes s_x \, \, ,
\label{H3D}
\end{equation}
where ${\cal M}({\bf k}) $=$M_0$-$B_2(k_x^2+k_y^2)$-$B_1 k_z^2$,  $\tau_x$, $\tau _y$ and $\tau_z$ are Pauli matrices acting on the parity of the states and $I$ is the unity matrix. 
In Hamiltonian (\ref{H3D}) we have neglected a ${\bf k}$-dependent diagonal term that only alters  the bands curvatures without modifying
the properties associated with topology. 
For a given  topological insulator of the Bi$_2$Se$_3$ family  the Hamiltonian parameters are 
obtained from {\it ab initio} electronic  calculations \cite{Liu_2010}. In the case of Bi$_2$Se$_3$
the ${\bf k} \cdot {\bf p}$  parameters are \cite{Zhang_2009,Liu_2010}
$M_0$=0.28 eV, $A_1$=2.2 eV\AA, $A_2$=4.1 eV\AA, $B_1$=10 eV\AA$^2$ and  $B_2$=56.6 eV\AA$^2$.
The spin operators written in the basis of Hamiltonian (\ref{H3D}) have the form \cite{Silvestrov_2012}
\begin{equation}
S_x=\tau_z \otimes  s _x \, , \, \, S_y= \tau _z \otimes s  _y \, , \, \, S_z = I \otimes s_z \,  .
     \label{3Dspin}
\end{equation}

Because of the non trivial topology of Hamiltonian (\ref{H3D}), surface states in the bulk gap will appear for any surface termination. The bands of these states are represented by Dirac-like Hamiltonians with the general form \cite{Liu_2010,Silvestrov_2012,PhysRevB.86.081303,Brey_2014},
\begin{equation}
H^{\nu} _{Surf} = (\boldsymbol{ \sigma}  \times \boldsymbol{ \kappa} )_{\nu} \, \, \, \, {\rm with} \,  \, \,  \boldsymbol{ \kappa} \equiv (A_2 k_x,A_2 k_y,A_1k_z) \, \, ,
\label{Hsurf}
\end{equation}
here $\nu$ indicates the direction normal to the surface. The matrices ${\boldsymbol \sigma }$ act on the basis of the surface Hamiltonian, that depends on the surface orientation \cite{Brey_2014},
and only coincides with the electron spin operator for  surfaces perpendicular to the $\hat z $-direction. 
For $z$-surfaces, an exchange field $\Delta$ directed in the $\nu$-direction introduces a term $\Delta \sigma _{\nu}$ in the surface Hamiltonian and therefore only a exchange field pointing in the $z$-direction open a gap at the Dirac point. 
The matching conditions between states of difference surfaces 
have the form 
\begin{equation}
\left ( \begin{array} {c}     c^{\nu}_u \\  c^{\nu}_v   \end{array} \right ) = {\Large M} _{\nu,\mu}
\left ( \begin{array} {c}     c^{\mu}_u \\  c^{\mu}_v  \end{array}\right ) \, \, ,
\label{Mxz}
\end{equation}
where $(c^{\nu}_u , c^{\nu}_v)$ is the spinor eigenstate of the Dirac  equation corresponding to surface $\nu$, and   explicit forms of the matrices ${\Large M} _{\nu,\mu}$ are given in reference \cite{Brey_2014}. For this work, the relevant connecting matrices are 
\begin{equation}
{\Large  M } _{z,x}  \!   \! = \! \!{\Large  M } _{\bar {z},\bar {x}} \!  \!  = \! \!  \sqrt {\frac {A_1}{2A_2} }\!
\left (  \! \! \begin{array} {cc} 1&  1\\  -1  &  1
  \end{array} \!  \! \right )  , 
{\Large  M } _{x,\bar {z}}   \!  \! = \! \! {\Large  M } _{\bar {x},z}  \!  \!  = \! \! 
{\Large  M } _{z,x} ^{-1} \;.
\label{M11}
\end{equation}  
\subsection{Tight-Binding Hamiltonian}
The band structure and topology of Hamiltonian (\ref{H3D}) can also be described by a tight-binding (TB) model. We consider a simple cubic crystal, such that each point of the lattice is characterized by 
an index ${\bf i}$=$(i_x,i_y,i_z)$. At each lattice point we associate four orbitals corresponding to positive and negative parity and spin up and down.  
We obtain the tight-binding hopping parameters, by considering the ${\bf k} \cdot {\bf p}$ Hamiltonian as a long wavelength expansion of the the 
tight-binding model, i.e.
$\sin (k a) \! \rightarrow \! ka $ and  $\cos(k a) \! \rightarrow \! 1 \! - \! \frac {(ka)^2} 2$,
$a$ being the lattice parameter. Finally, for the cubic lattice, we relate $\cos{k _{\nu} a}$ and $\sin{k _{\nu}a}$  with  the sum  and subtraction respectively, of the Bloch phases  between a site and its two  first neighbors in the $\nu$-direction. Therefore  the cosine  is related with a real hopping parameter  and the sine  with a pure imaginary hopping. 
Taking these considerations into account, the tight-binding Hamiltonian reads,
\begin{widetext}
\begin{eqnarray}
H^{TB} & = & \sum_{\bf i} \left\{ M_0 \,  {\bf c} ^+ _{\bf i} \tau _z \otimes I  \, {\bf c} _i   -   \frac {B_1} {a ^2} \left ( ({\bf c} ^+ _{\bf i} - {\bf c} ^+ _{{\bf i} +\hat z}) \, \tau_z \otimes I \, 
({\bf c}  _{\bf i} - {\bf c}  _{{\bf i} +\hat z}) \right ) \right. \nonumber \\
 &-&  \frac {B_2} {a ^2} \left ( ({\bf c} ^+ _{\bf i} - {\bf c} ^+ _{{\bf i} +\hat x})  \, \tau_z \otimes I \, ({\bf c}  _{\bf i} - {\bf c}  _{{\bf i} +\hat x}) +({\bf c} ^+ _{\bf i} - {\bf c} ^+ _{{\bf i} +\hat y}) \, \tau_z \otimes I \, ({\bf c}  _{\bf i} - {\bf c}  _{{\bf i} +\hat y})
\right )\nonumber \\
&-&  \frac {A_1} {2 a}i \left ( {\bf c} ^+ _{\bf i} \, \tau _x \otimes s_z \,  {\bf c} _{{\bf i}+ \hat z} -  {\bf c} ^+ _{{\bf i}+\hat z} \, \tau _x \otimes s_z \, {\bf c} _{\bf i} \right )  \nonumber \\
&-&  \left. \frac {A_2} {2 a}i \left ( {\bf c} ^+ _{\bf i} \, \tau _x \otimes s_x \, {\bf c} _{{\bf i}+ \hat x} -  {\bf c} ^+ _{{\bf i}+\hat x} \, \tau _x \otimes s_x \, {\bf c} _{\bf i} 
+ {\bf c} ^+ _{\bf i} \, \tau _y \otimes \, s_x \, {\bf c} _{{\bf i}+ \hat y} -  {\bf c} ^+ _{{\bf i}+\hat y} \, \tau _y \otimes s_x \,  {\bf c} _{\bf i} \right ) \right\}  \; ,
\label{HTB}
\end{eqnarray}
\end{widetext}
where ${\bf c _i}$=$(c_{{\bf i} , +,\uparrow},-i c_{ {\bf i} , -,\uparrow},c_{{\bf i} , +,\downarrow}, i c_{ {\bf i} , -,\downarrow})$, and the operator $c_{{\bf i},\tau,s}$ annihilates an electron at site ${\bf i}$ with parity $\tau$ and spin $s$.

\begin{figure}[h!]
\begin{minipage}{1.0\linewidth}
 \includegraphics[width=1.0\textwidth]{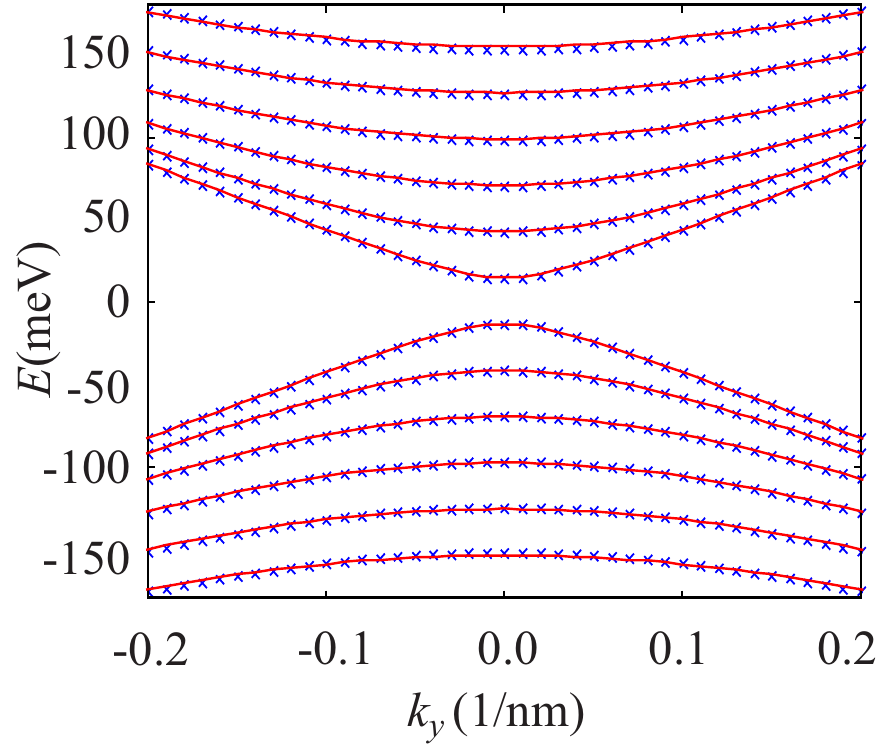}
\end{minipage}
 \caption{(Color online) Band structure of a quantum wire along the $\hat y$-direction of Bi$_2$Se$_3$ obtained from  solving the TB Hamiltonian (\ref{HTB}) (crosses) and
by matching surface states, Eq. (\ref{enky}). The dimensions of the wire are $N_x$=$N_z$=15 and the lattice parameter of the lattice is $a$=1 nm. }
 \label{BandasWire}
\end{figure}

In order to check the adequacy of the tight-binding Hamiltonian for describing the low energy properties of TI, we calculate the band structure of a TI quantum wire with rectangular cross section.  The wire has $N_x$ and $N_z$ sites along the $x$ and $z$ axes, respectively, and we take it infinitely long in the $\hat y$-direction, so that $k_y$ is a good quantum number.   In Fig. \ref{BandasWire}, we plot the band structure  obtained by diagonalizing the tight-binding Hamiltonian with the parameters corresponding to Bi$_2$Se$_3$ and a lattice parameter $a$=1 nm. In Bi$_2$Se$_3$ the bulk energy gap is 0.56eV, therefore the bands shown in Fig. \ref{BandasWire} should correspond to midgap surface states. We verify this by computing the energies of the surface states of the quantum wire using the effective Hamiltonians of the different surfaces of the wire, Eq. (\ref{Hsurf}), and the matching conditions Eqs. (\ref{Mxz}) and (\ref{M11}).  For this quantum wire geometry the surface states have an analytical expression \cite{Brey_2014},
\begin{equation}
\epsilon _{n,k_y} = \pm \sqrt{(A_2 k_y)^2+\left (    \pi \frac {A_1 A_2}{A_1 L_z+ A_2 L_x} (n - \frac 1 2 )\right ) ^2 } ,
\label{enky}
\end{equation}
with $n$=1,2,3... Analogous expression have been obtained for TI wires with cylindrical geometry \cite{Egger_2010,Kundu_2011}. In Fig. \ref{BandasWire} we plot the  quantum wire surface band structure, Eq. (\ref{enky}), as function of the wavenumber $k_y$
for $L_x$=$L_z$=$\left(N_x+1\right)a$. The agreement between both calculations indicate that
the tight-binding Hamiltonian (\ref{HTB}) describes very accurately the surface states of a nanowire.

\subsection{Conductance}
For computing the conductance of a two terminal geometry like the one depicted in Fig. \ref{TI-junction}, described by a tight-binding Hamiltonian, we use a recursive Green function method. The rectangular wire
is defined along the $\hat{y}$ direction and connected to normal metallic leads at both ends.
The lateral dimensions are $N_x$=$N_z$ sites, and the number of layers along the wires is $N_y=2N_c+1$, with the first  and last layers corresponding to the metallic leads. These leads are described by constant imaginary self-energies $\Sigma^{a,r}_{L,R} = \pm i \gamma P_{L,R}$, where $P_{L,R}$ are projectors on the sites of the first and last layer respectively. 
The energy dependent conductance of the system is \cite{Fisher_1981}
\begin{equation}
G(E)=\frac{4e^2}{h} {\rm Tr} \left [\hat { \Gamma }_L     \,  \hat {\cal G} ^r _{LR} (E) \, \hat { \Gamma }_R   \,   \hat {\cal G} ^a _{RL} (E)   \right ]
\end{equation}
where $\hat { \Gamma }_{L,R}=\mbox{Im} \Sigma^a_{L,R}$ and
$\hat {\cal G} ^{r,a} _{LR} (E)$ are the $16 N_x N_z\times 16 N_x N_z$ matrix elements of the
Green function operator,
\begin{equation}
{\cal G}^{a,r} (E) =(E-H^{TB}-\Sigma^{a,r} _L -\Sigma ^{a,r}_R) ^{-1},
\end{equation}
connecting the outermost layers of the wire. To calculate these quantities we use conventional recursive  Green's  function techniques (see e.g. \cite{Yeyati_1992,Lewenkopf_2013,Thorgilsson_2014}).

\begin{figure}[h!]
\begin{minipage}{1.0\linewidth}
 \includegraphics[width=1.0\textwidth]{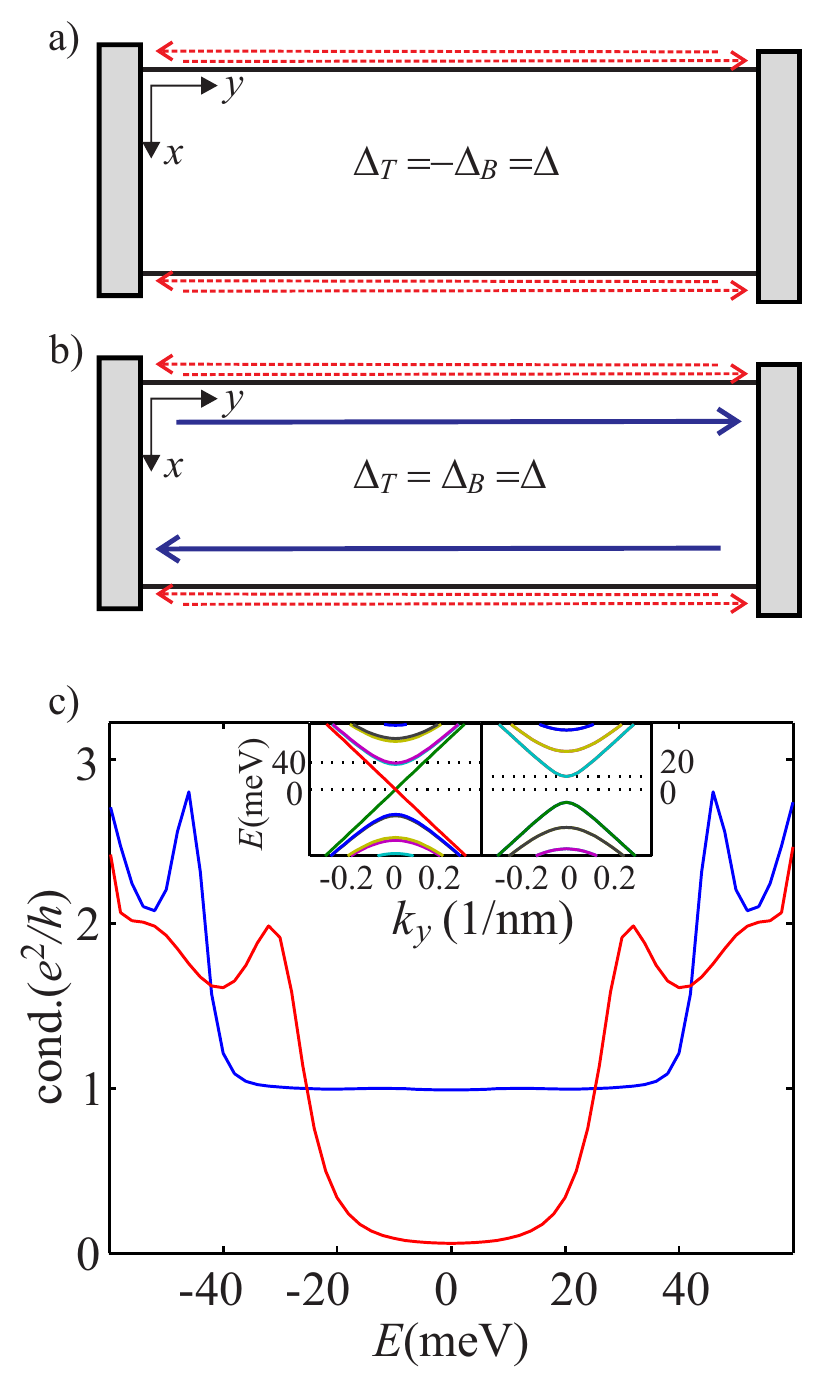}
\end{minipage}
 \caption{(Color online) Schematic representation of the edge  charge  carrying channels in the 
(a) $(\Delta \bar \Delta\Delta\bar \Delta)$ and (b) $(\Delta \Delta\Delta\Delta)$ configurations. The red dashed lines represent 
non-chiral states confined on the laterals $\pm x$-surfaces. In (b) the continuum blue lines indicate chiral edge states. 
In (c) we plot the conductance as function of energy for the $(\Delta \Delta\Delta\Delta)$ (blue) and $(\Delta \bar \Delta\Delta\bar \Delta)$ (red) configurations.
The peak in the conductance at  $\sim$45 meV  ($\sim$ 32 meV) in the equal (opposite) magnetization cases corresponds to the lowest non-chiral energy confined state on the laterals surfaces.
The insets in (c) show
the band structure for the $(\Delta \Delta\Delta\Delta)$ (left panel), and $(\Delta \bar \Delta\Delta\bar \Delta)$ (right panel) cases. The dimensions of the wire are $N_x$=$N_z$=15 and $N_y$=59. The lattice parameter is $a$=1 nm and the exchange field $\Delta$= 90 meV. }
 \label{Fig++++}
\end{figure}

\section{Results and Discussion}
\subsection{Uniform Magnetization}
We first consider  systems with uniform magnetization on top and bottom  surfaces, i.e. geometries with exchange fields $(\Delta _T\Delta_B\Delta_T\Delta_B)$. In the tight-binding formalism, the  exchange fields acts as a Zeeman coupling on the lattice surface sites.  In these calculations we fix the  lateral dimensions of the wires to $N_x$=$N_z$=15, and the length of the wires is $N_y$=59. We take a lattice parameters $a$=1 nm,  large enough or avoid the direct coupling between surface states of opposite surfaces. In all cases we fix the magnitude of the exchange field to 90 meV.

In Fig. \ref{Fig++++} we plot the results for the equal $(\Delta \Delta\Delta\Delta)$ and opposite $(\Delta \bar \Delta  \Delta \bar\Delta)$ configuration cases.
In both cases the $z$-surfaces are gapped and should have a half-integer anomalous Hall conductivity. 
In the equal magnetization case, there are mid gap energy chiral edge states, see inset of  Fig.  \ref{Fig++++}c, with  wavefunctions that decay exponentially in the $z$-surfaces and are constant in the
lateral $x$-surfaces \cite{Brey_2014}. Mathematically these solutions 
appear when matching the surface states determined by Eq. (\ref{Hsurf}) for $x$ and $z$ surfaces
using the connecting matrices of Eq. (\ref{M11}). Physically, the existence of  the chiral edge states indicates that the wire behaves as an integer anomalous quantum Hall system. In the opposite magnetization case, it is not mathematically possible to match the mid gap states in top and bottom surfaces 
through the lateral $x$-surface chiral states, and the wire as a whole behaves as an normal insulator. These results indicate that the $(\Delta \bar \Delta  \Delta \bar\Delta)$ configuration preserves time reversal symmetry.  In the equal polarization configuration the existence of chiral  edge states quantizes 
the conductance of the system to  $e^2/h$, Fig. \ref{Fig++++}c. On the contrary,  when the exchange field has opposite sign on opposite surfaces the lack of chiral edge states only a residual conductance
$G(E) \ll e^2/h$ due to finite size effects appears at low energies. 
In both configurations, for larger absolute values of the chemical potential, higher energy lateral confined states yield an extra 2$e^2/h$ contribution to the conductance.  

\begin{figure}[h!]
\begin{minipage}{1.0\linewidth}
 \includegraphics[width=1.0\textwidth]{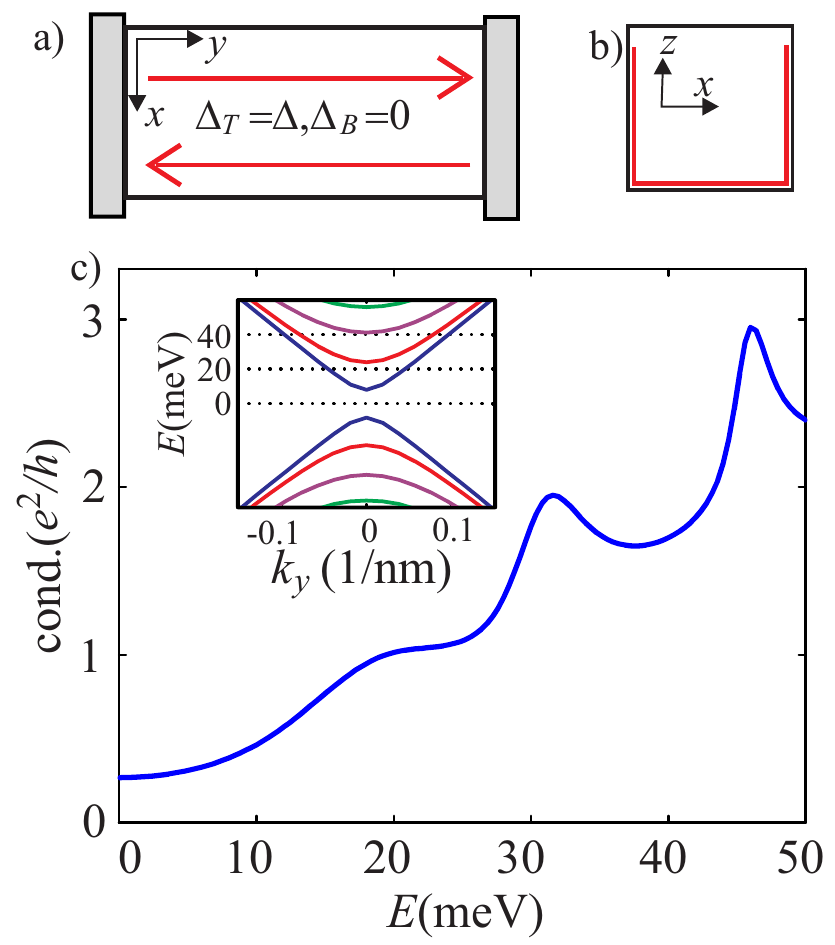}
\end{minipage}
 \caption{(Color online) (a) Schematic representation of the charge carrying channels in the 
$(\Delta 0 \Delta 0)$ configuration. The red lines represent the lowest energy
states confined along the $\pm x$-surfaces and -$z$-surfaces. This state is not chiral and is spin 
polarized. In (b) the red lines around the surface on a $x-z$ cross section indicate the region where
these states have a significant weight. Finally, in (c) we plot the conductance  as function of energy for this configuration. 
The inset in (c) shows
the corresponding band structure for an infinite wire. Due to the exchange field in the top layer, the lowest energy confined states are spin polarized.
The wire dimensions and other parameters as in Fig. \ref{Fig++++}. } 
 \label{Fig+0+0}
\end{figure}

In Fig. \ref{Fig+0+0} we show the conductance for a TI wire with an exchange field applied uniformly just on the +$z$-surface, i.e for a  $(\Delta 0 \Delta 0)$ configuration. In this geometry only the top surface has a half-integer quantized Hall conductivity. In absence of electron-electron interaction
chiral edge states carry an integer charge, and therefore it can not be possible to observe a half-integer QHE. This is consistent with the impossibility of
matching the mid gap energy solutions of the surface Hamiltonians (\ref{Hsurf}) corresponding to this configuration. The lowest energy confined state is  non chiral and corresponds to a wave function confined along the bottom and lateral surfaces (see Fig. \ref{Fig+0+0}b).
The exchange field on the top surface makes the low energy confined states spin polarized, 
this results in a shoulder in the conductivity at  $e^2/h$ and peaks at $2e^2/h$ and $3e^2/h$ as a function of energy.

\begin{figure}[h!]
\begin{minipage}{1.0\linewidth}
 \includegraphics[width=1.0\textwidth]{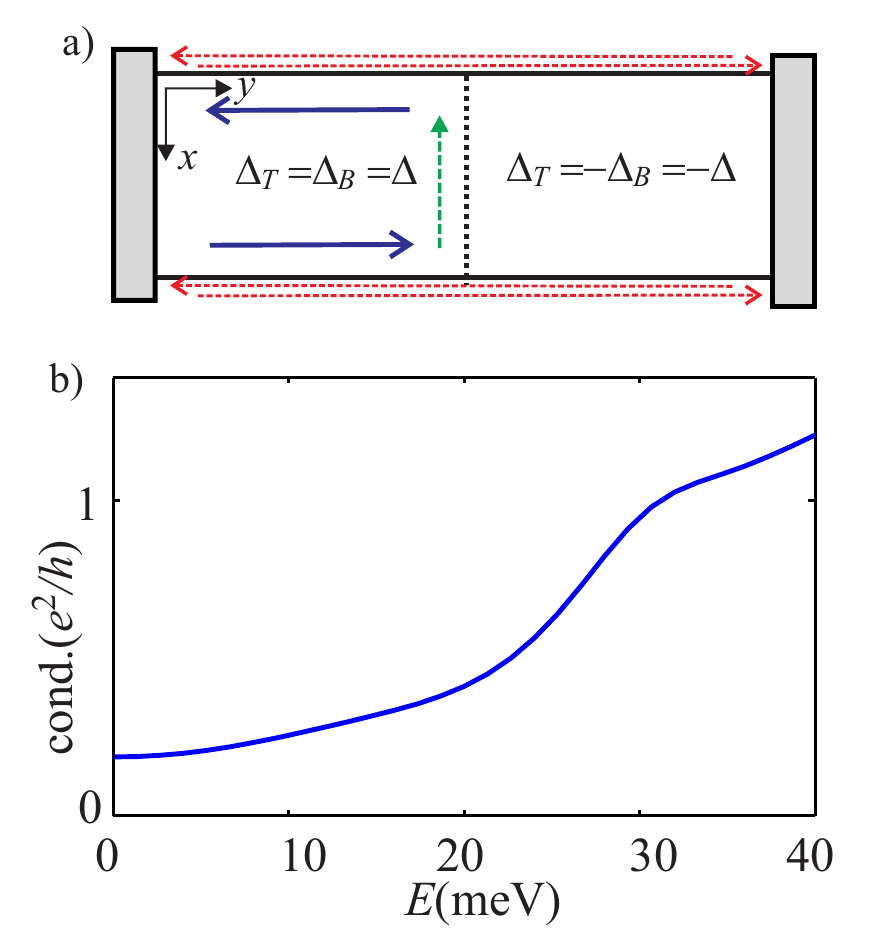}
\end{minipage}
 \caption{(Color online) (a) Schematic representation of the charge carrying channels of a TI wire in the $(\Delta\Delta\bar \Delta\Delta)$ configuration.
The red dashed lines  represent non chiral  states confined  along the  $\pm x$-surfaces. 
The continuum blue lines indicate lateral chiral edge states. The green dashed line represents a chiral bound mode separating regions with opposite magnetization in the top surface.
In (b) we plot the conductance as function of energy.  The wire dimensions and other parameters as in Fig. \ref{Fig++++}.}
 \label{Fig++-+}
\end{figure}

\subsection{Non Uniform Magnetization}
In this subsection we consider geometries where the magnetization in one or in both $z$-surfaces is not uniform. 
When joining two surfaces with different quantum Hall conductivities some chiral bound modes should appear at the junction \cite{Martin_2008,hasan_2010}.  Similar states have been predicted to appear at 
p-n junctions in the presence of a magnetic field \cite{PhysRevB.77.081404,PhysRevB.85.235131}.The number of chiral modes is equal to the change in Chern number when traversing the junction. In two dimensions this is equivalent to the difference in Hall conductivities in units of $e^2/h$. The charge current  direction of this chiral bound modes should be the appropriated to preserving current conservation \cite{hasan_2010,Beenakker-Book}. 
In a magnetically doped $z$-surface the Hall conductivity is quantized to $\pm e^2/2h$, with the sign depending on the orientation of the magnetization. Therefore, at the junction between two regions of a $z$-surface with opposite exchange field, a single chiral bound mode should appear. 

In Fig. \ref{Fig++-+} we plot the conductance for a wire  with the bottom surface  uniformly magnetized in the $z$-direction whereas the left and right parts of the top surface are polarized in the positive and negative direction respectively, i.e. a
$(\Delta\Delta\bar \Delta\Delta)$ configuration. In this geometry an edge channel appears at the middle of the top surface. At low energy, this state produces a perfect backscattering of the current carried by the chiral state confined in the $x$-surface through the chiral edge state in the -$x$-surface. The shoulder at $e^2/h$ that appears in the conductance for energies near 30 meV corresponds to the coupling between the incoming chiral edge mode with the non-chiral confined state in the $(\Delta\bar{\Delta}\Delta\bar{\Delta})$ configuration of the right part of the wire, see Fig. \ref{Fig++++}.

\begin{figure}[h!]
\begin{minipage}{1.0\linewidth}
 \includegraphics[width=1.0\textwidth]{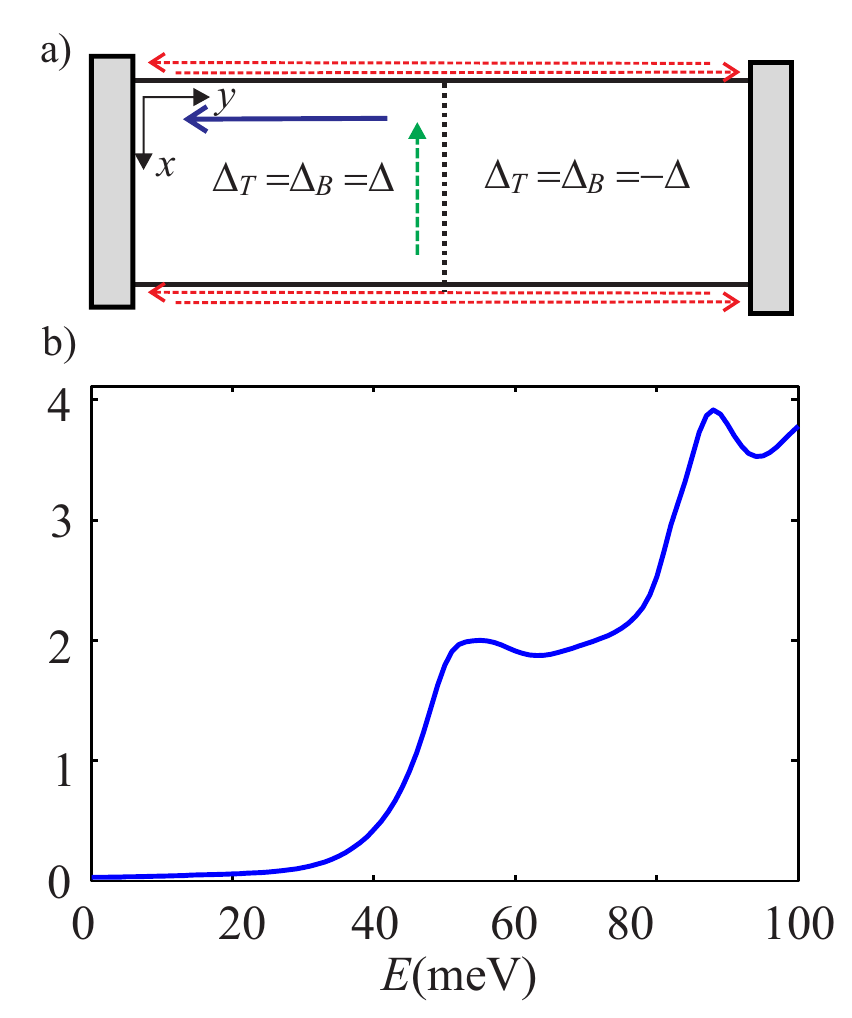}
\end{minipage}
 \caption{(Color online)  a) Schematic representation of the charge carrying channels of a TI wire in the $(\Delta\Delta \bar\Delta \bar\Delta)$ configuration.
The red dashed lines  represent non chiral  states confined  along the   $\pm x$-surfaces. 
The continuum blue lines  indicate lateral chiral edge states. The green dashed (continuous) line represents a chiral bound mode separating regions  with opposite magnetization in the top (bottom) surface.
In (b)we plot the conductance  as function of the Fermi energy of the  wire.  The wire dimensions
and other parameters as in Fig. \ref{Fig++++}. }
 \label{Fig++--}
\end{figure}

Chiral bound states also occur when the left and right parts of the TI wire have opposite sign Hall conductivity.  In Fig. \ref{Fig++--}b we plot the conductance for the ($\Delta\Delta\bar \Delta \bar \Delta)$ configuration. Here the top and bottom left surfaces has a positive exchange field and globally a positive integer quantum Hall conductivity.  In the right part of the wire the orientation of the exchange fields and accordingly the sign of the Hall conductivity are reversed.  As a consequence, at the junction between left and right parts, two chiral edge modes, carrying charge in the same direction appear at the junction. One of these modes is located in the top surface whereas the other is located in the opposite one.  The chiral bound states reflect the incident current perfectly and at low energy the conductance is negligible.  At $E \sim 55$ meV the conductance jumps to $2e^2/h$ due to the contribution of the lateral confined non-chiral state, in agreement with the band structure for the equal magnetization case illustrated in Fig. \ref{Fig++++}.

\begin{figure}[h!]
\begin{minipage}{1.0\linewidth}
 \includegraphics[width=1.0\textwidth]{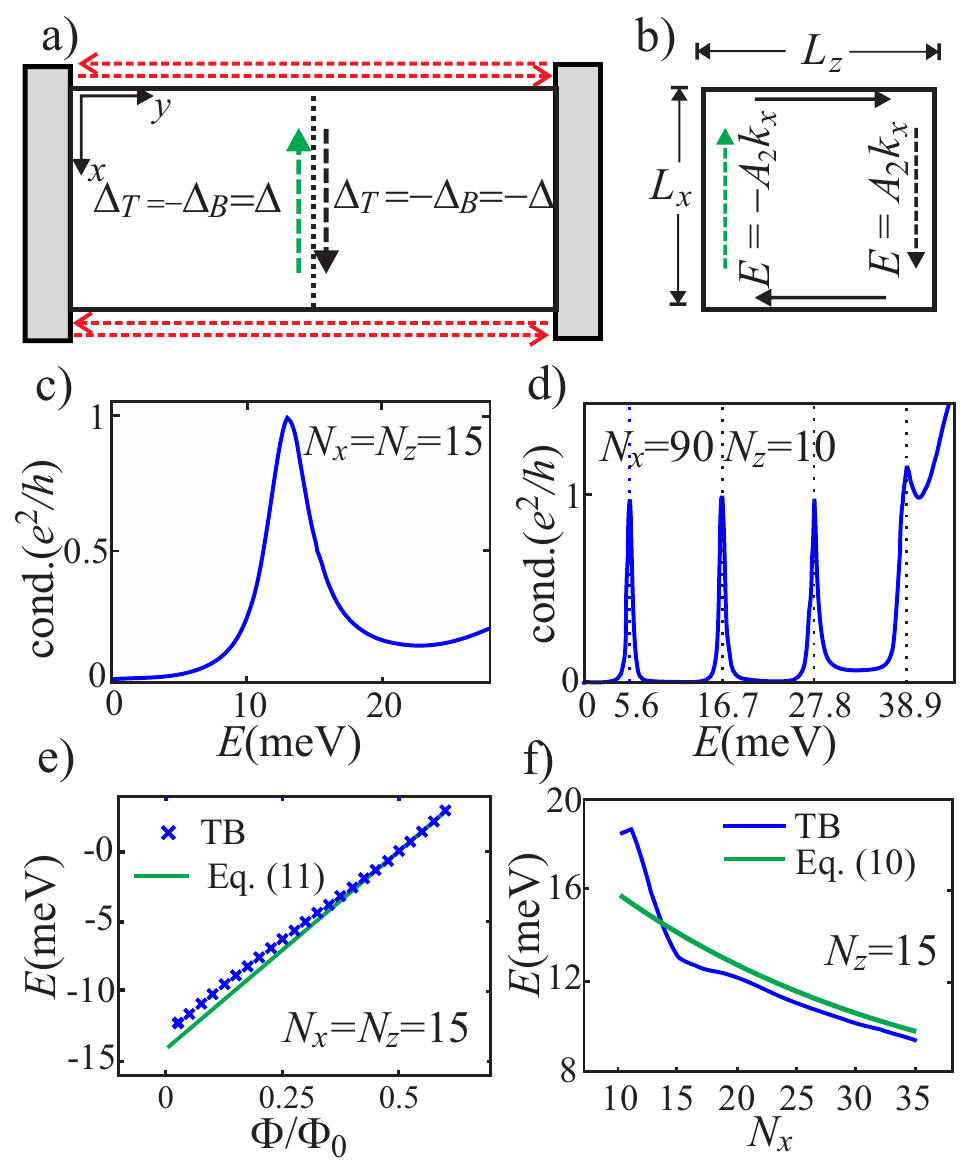}
\end{minipage}
 \caption{(Color online) a) and b)  Schematic representation of the charge carrying channels of a TI wire in the $(\Delta \bar\Delta \bar\Delta \Delta)$ configuration.
The red dashed lines  represent non chiral  states confined  along the   $\pm x$-surfaces. The green and black dashed lines indicate chiral edge  bound states
in the $\pm$-surfaces. In b) The continuous blue lines represent chiral edge bound states in the $\pm x$-surfaces.
The continuum blue lines  indicate lateral chiral edge states. The green dashed (continuous) line represents a chiral bound mode separating regions  with opposite magnetization in the top (bottom) surface. In c) we plot the conductance as function of energy for $N_x$=$N_z$=15 and $N_y$=59. d) conductance as function of energy for  $N_x$=90, $N_z$=15 and $N_y$=59.
e) Energy of the lowest confined state as function of a magnetic flux applied parallel to the $\hat y$. In f) we plot the energy of the lowest confined state, obtained from the tight-binding calculation and from Eq. (\ref{level}) as function of $N_x$ for $N_z$=15.  All other parameters as in Fig. \ref{Fig++++}.
}
 \label{Scheme+--+}
\end{figure}

\subsection{Bound states in the antisymmetric ($\Delta \bar \Delta \bar \Delta \Delta$) configuration} 

An interesting situation occurs when the chiral edge bound states on top and bottom surfaces have opposite velocities. This case happens in the ($\Delta \bar \Delta \bar \Delta \Delta$) geometry, where the exchange fields change sign when going from left to right and from bottom to top of the wire. In this configuration, at the junction between left and right parts of the 
$\pm z$-surfaces there appear opposite directed chiral bound states, see Fig. \ref{Scheme+--+}. 
Because the magnetization on top and bottom layers has opposite directions there are no chiral states in the lateral $\pm x$-surfaces. 
However, in this geometry there are chiral bound states at the junction between the left and the right parts of the wire. The wavefunctions in the $\pm x$-surfaces are confined by the exchange gaps in the 
$\pm z$-surfaces. The magnetization has opposite sign on the left and right parts of the TI wire, and therefore the wavefunction corresponding to the lowest positive (highest negative) energy state in the right (left) part of the $x$-surface coincides with the wavefunction of the highest negative (lowest positive) energy state in the left (right) part of the sample. This crossing of the positive and negative energy states when crossing the junction between left and right parts, justify the existence of a chiral bound mode in the $x$-surface. With the same arguments we expect a chiral bound state in the -$x$-surface pointing in the opposite direction.

In Fig. \ref{Scheme+--+}c we plot the conductance as function of energy for a wire with magnetization ($\Delta \bar \Delta \bar \Delta \Delta$) and $N_x$=$N_z$=15 and $N_y$=59. There is a well defined peak with a maximum value of $e^2/h$ that indicates resonant tunneling through a confined state.  When the perimeter of the wire increases, the energy of the confined state decreases and more confined states appear in the gap energy, see Fig. \ref{Scheme+--+}d.  In addition the energy of the lowest confined states is half the energy separation between higher confined states. This suggests that these states embrace the wire and are formed by matching the four one dimensional bound chiral modes in the 
$\pm x$ and $\pm z$ surfaces \cite{Brey_2014}. 

The wave function of these chiral confined modes in the $\pm z$-surfaces can be approximately obtained
by using the  surface Hamiltonians, Eqs. (\ref{Hsurf}). For the $\pm z$-surfaces the wave function have the form $\psi ^{\pm z} (x,y)$=$ (1,-i) ^{T }e ^{\pm ik_x x} e^{-\Delta |y|}$, with $k_x >0$.  These states are bound near $y$=$0$ and have opposite velocities on opposite surfaces. Their spin is, however, well defined along the $y$-direction. An adequate description of the chiral bound modes in the $\pm x$-surfaces requires a basis of electronic states larger then the used with the surface Hamiltonians. However, an approximate ansatz for these wavefunctions is provided by, $\psi ^{\pm x} (y,z)$=$ (i,1) ^{T }e ^{\mp ik_z z} e^{- \tilde{\Delta} |y|}$, with $k_z >0$ and $\tilde {\Delta}$ the effective gap in the $\pm$-surfaces.
When an electron moves enclosing  a loop around the TI wire, the matching of the wave function, 
Eqs. (\ref{Mxz}-\ref{M11}) gives the quantization condition,   
\begin{equation}
E_{n}=\frac{A_{1}A_{2}\left( n+\frac{1}{2} \right)
\pi }{A_{2}L_{z}+A_{1}L_{x}}\;,n=0,1,...  \label{level}
\end{equation}
The $\pi/2$ that appear in the quantization energy, occurs due to the helical nature of the carriers, when the electrons wind around the TI wire, the expectation value of the Pauli matrices that appear in the surface Hamiltonians rotates by $2\pi$, and the electron wave function acquires a phase  of $\pi$ \cite{Luo_2009}. This shift in the  quantization condition agrees perfectly with the fact that 
the energy of the lowest confined state is half the separation in energy of other confined states. 
In Fig. \ref{Scheme+--+}e we plot the energy of the confined states as function of the lateral size of the wire $N_x$ for $N_z$=15. The agreement between the results obtained with the tight-binding calculations and Eq. (\ref{level}) is good at larger values of $N_x$, this reflects the fact that there is a gap in the $\pm x$ surfaces induced by the the top and bottom surfaces. 

Finally, to check that the confined states correspond to a wave function encircling the wire, we have added a magnetic flux, $\Phi$, through the wire cross section. The bound states should evolve according to
\begin{equation}
E_{n}(\Phi)=\frac{A_{1}A_{2}\left( n+\frac{1}{2}  + \frac{\Phi}{\Phi_0} \right)
\pi }{A_{2}L_{z}+A_{1}L_{x}} \;,
\label{level-flux}
\end{equation}
where $\Phi_0$ is the flux quantum.
In Fig. \ref{Scheme+--+}e we plot the energy of the first confined state obtained from tight-binding calculations as function of the magnetic flux.  As can be observed, the energy of this resonant level  follows closely the analytical prediction of Eq. (\ref{level-flux}) and vanishes as expected at $\Phi = \Phi_0/2$. 

\

\section{Conclusions}
In this work we have analyzed the electronic structure and the conductance of a TI nanowire with uniform and non-uniform magnetization on its surfaces. For the uniform case with equal polarization on the $\pm z$-surfaces, conduction through chiral states propagating along the $\pm x$-surfaces yields a quantized $e^2/h$ conductance at low energies. The conductance could be almost fully suppressed by reversing the magnetization on one of the surfaces or on one half of the wire, thus providing a switching mechanism. On the other hand, for the non-uniform case in the anti-symmetric $(\Delta \bar\Delta \bar\Delta \Delta)$ configuration, chiral bound states appear at the center of the wire, leading to resonant tunneling with a conductance raising up to $e^2/h$ for symmetric coupling to the left-right leads. The energy of these states is well described by a simple analytical expression, showing that it depends on the wire perimeter and can be controlled by a magnetic flux through its cross section. The magnetically doped nanowires thus provide a physical realization of a highly tunable spin-polarized quantum dot.

\section{acknowledgments}
The authors would like to thank R. Egger for useful comments on the manuscript.
Funding for this work was provided by Spanish MINECO
through grants FIS2012-33521 and FIS2014-55486 and COLCIENCIAS through grant 110165843163.


\end{document}